\documentclass[sigconf]{acmart}

\AtBeginDocument{%
  \providecommand\BibTeX{{%
    \normalfont B\kern-0.5em{\scshape i\kern-0.25em b}\kern-0.8em\TeX}}}

\copyrightyear{2024}
\acmYear{2024}
\setcopyright{acmlicensed}\acmConference[WWW '24 Companion]{Companion Proceedings of the ACM Web Conference 2024}{May 13--17, 2024}{Singapore, Singapore}
\acmBooktitle{Companion Proceedings of the ACM Web Conference 2024 (WWW '24 Companion), May 13--17, 2024, Singapore, Singapore}
\acmDOI{10.1145/3589335.3648340}
\acmISBN{979-8-4007-0172-6/24/05}

\usepackage{threeparttable}
\usepackage[utf8]{inputenc} 

\usepackage{booktabs}       
\usepackage{nicefrac}       
\usepackage{microtype}      
\usepackage{xcolor}         
\usepackage{soul}
\usepackage{url}
\usepackage[font={small,bf}]{caption}
\usepackage{graphicx}
\usepackage{amsmath}
\usepackage{amsthm}
\usepackage{algorithm}
\usepackage{algorithmic}
\usepackage{multirow}
\usepackage{subfigure}
\usepackage{bm}
\usepackage{color}

\usepackage{tablefootnote}

\usepackage{xspace}
\newcommand{\ModelName}{TIN\xspace}

\usepackage{stfloats}
\usepackage{marvosym}

\usepackage{pifont}
\newcommand{\cmark}{\ding{51}}%
\newcommand{\xmark}{\ding{55}}%

\usepackage{color, colortbl}
\definecolor{LightCyan}{rgb}{0.88,1,1}
\definecolor{color2}{rgb}{1,0.88,1}
\definecolor{color3}{rgb}{1,1,0.88}

\begin{document}

\title{Temporal Interest Network for User Response Prediction}


 \author{Haolin Zhou}\authornote{Equal Contribution}
 \affiliation{%
    \institution{
    Shanghai Jiao Tong University
    \country{}
    }}
 \email{koziello@sjtu.edu.cn}

\author{Junwei Pan}\authornotemark[1]
 \affiliation{%
    \institution{Tencent
    \country{}
    }}
 \email{jonaspan@tencent.com}

 \author{Xinyi Zhou}
 \affiliation{%
    \institution{
    Shanghai Jiao Tong University
    \country{}
    }}
 \email{zhouxy1003@sjtu.edu.cn}

 \author{Xihua Chen}
 \affiliation{%
    \institution{Tencent
    \country{}
    }}
 \email{tinychen@tencent.com}

  \author{Jie Jiang}
 \affiliation{%
    \institution{Tencent
    \country{}
    }}
 \email{zeus@tencent.com}

  \author{Xiaofeng Gao(\Letter)}
 \affiliation{%
    \institution{
    Shanghai Jiao Tong University
    \country{}
    }}
 \email{gao-xf@cs.sjtu.edu.cn}

   \author{Guihai Chen}
 \affiliation{%
    \institution{
    Shanghai Jiao Tong University
    \country{}
    }}
 \email{gchen@cs.sjtu.edu.cn}

\renewcommand{\shortauthors}{Haolin Zhou et al.}

\begin{abstract}

User behaviors are among the most critical features for user response prediction in recommendation systems.
Many works have revealed that a user's behavior reflects her interest in the candidate item, owing to their semantic or temporal correlation.
While the literature has individually examined each of these correlations, researchers have yet to analyze them in combination, that is, the semantic-temporal correlation. 
We empirically measure this correlation and observe intuitive yet robust patterns. 
We then examine several popular user interest models and find that, surprisingly, none of them learn such correlation well.

To fill this gap, we propose a Temporal Interest Network (TIN) to capture the semantic-temporal correlation simultaneously between behaviors and the target. 
We achieve this by incorporating target-aware temporal encoding, in addition to semantic encoding, to represent behaviors and the target. 
Furthermore, we conduct explicit 4-way interaction by deploying target-aware attention and target-aware representation to capture both semantic and temporal correlation. 
We conduct comprehensive evaluations on two popular public datasets, and our proposed TIN outperforms the best-performing baselines by 0.43\% and 0.29\% on GAUC, respectively. 
During online A/B testing in Tencent's advertising platform, TIN achieves 1.65\% cost lift, and 1.93\% GMV lift over the base model.
It has been successfully deployed in production since October 2023, serving the WeChat Moments traffic. 
We have released our code at \url{https://github.com/zhouxy1003/TIN}.

\end{abstract}

\ccsdesc[500]{Information systems~Display advertising}
\ccsdesc[500]{Computing methodologies~Neural networks}
\ccsdesc[500]{Computing methodologies~Factorization methods}

\keywords{User Response Prediction, Target Attention, Sequential Recommendation, CTR Prediction}
    
    
    \maketitle

    \section{Introduction}

In recent decades, users have been confronted with overwhelming information while browsing websites or mobile apps. 
This poses significant challenges for electronic retailers, content providers, and online advertising platforms in their quest to recommend suitable items to individual users within specific contexts. 
Consequently, recommendation systems have gained widespread deployment to capture users' interests and predict their preferences from an extensive pool of candidate items.
For instance, in cost-per-click (CPC) advertising, advertising platforms must bid for each ad based on the estimated value of the impression, which relies on the bid value and the estimated Click-Through Rate (CTR). 
As a result, the accurate prediction of user response becomes a critical factor and has garnered considerable research attention.

\begin{figure}[ht!]
    \centering
    \captionsetup{labelfont=bf}
    
    \subfigure[\centering Ground truth STC]{
        \begin{minipage}[b]{0.45\linewidth}
    	\label{subfig:674}
  		\includegraphics[width=1\linewidth]{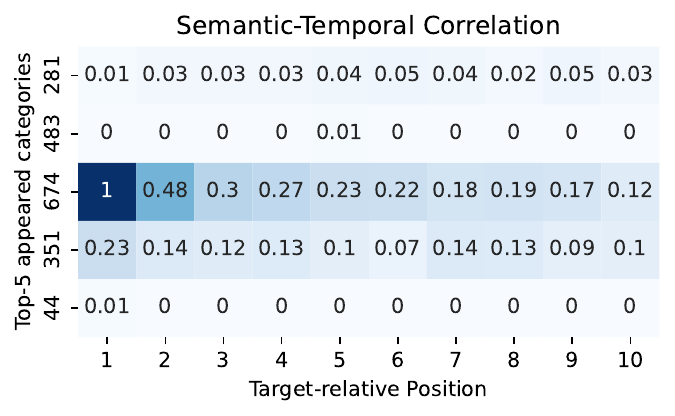}
        \end{minipage}}
    \subfigure[\centering DIN's learned STC (0)]{
		\begin{minipage}[b]{0.45\linewidth}
			\label{subfig:din_plus_674}
   \includegraphics[width=1\linewidth]{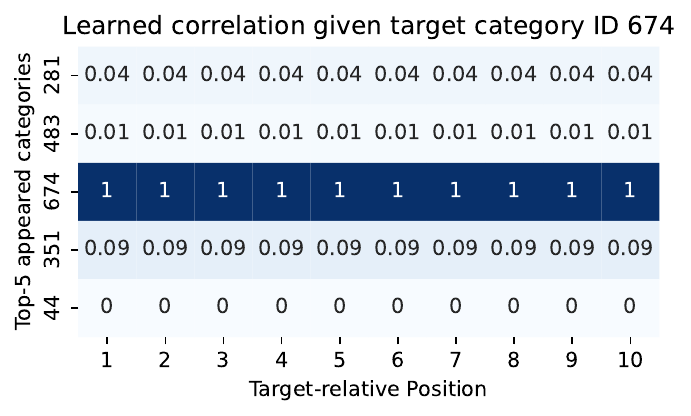}
	\end{minipage}}
	\subfigure[\centering SASRec's learned STC (0.8461)]{
		\begin{minipage}[b]{0.45\linewidth}
			\label{subfig:sas_674}
   \includegraphics[width=1\linewidth]{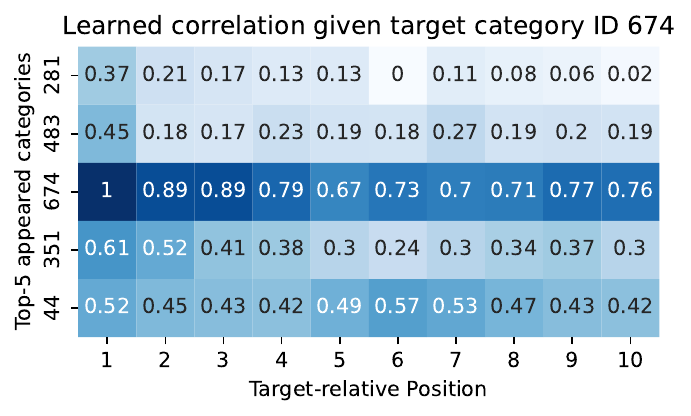}
	\end{minipage}}
	\subfigure[\centering BST's learned STC (0.6245)]{
		\begin{minipage}[b]{0.45\linewidth}
			\label{subfig:bst_674}
 	\includegraphics[width=1\linewidth]{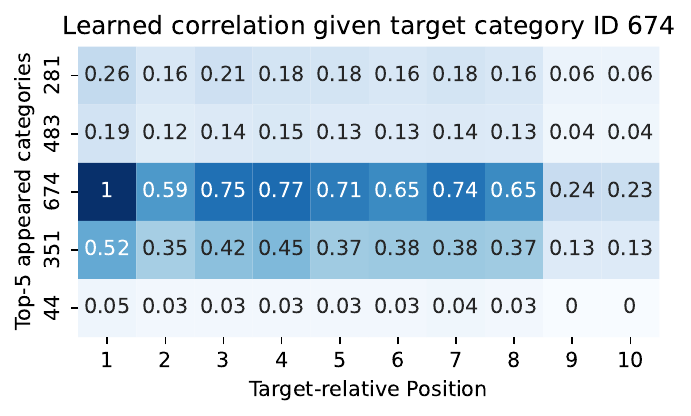}
	\end{minipage}}
    \caption{(a): The ground truth Semantic-Temporal Correlation (STC) between history behaviors of Top-5 frequent categories (y-axis) at various target-relative positions from 1 to 10 (x-axis), and the target category of \texttt{674}; (b,c,d): the learned STC in DIN, SASRec and BST. Numbers in parentheses denote the Pearson correlation coefficient between ground truth STC and learned STC on behaviors with the same category as the target (3rd row in each figure), \textit{i.e.}, \texttt{674}.}
    \label{fig:ground_truth_and_learned_qc}
\end{figure}

Recommendation systems usually use categorical ID features to represent users, items, and the context, such as user demographic features, item attribute features, \textit{etc}.
Among them, users' history behaviors are crucial since they represent a user's preference over items in the past and, therefore, may reflect her interest in the candidate item.
Many works tried to capture such user interest by recurrent neural networks~\cite{dien2019, GRU4Rec}, self-attention~\cite{SASRec, Bert4Rec, BST}, or target attention~\cite{din2018, dien2019, dsin2019, BST}. 


Intuitively, a user's interest is reflected by what items she has interacted with and when these interactions happened.
For example, if she has clicked ads on video games many times, then it's a good idea to recommend ads for new video games to her in the future. 
However, if all her clicks on video games happened 1 year ago, but she clicked on clothing ads several times recently, then it's better to recommend clothing ads.
Therefore, when capturing a user's interest in a candidate item, it's crucial to consider both the semantic and temporal correlations between behaviors and the candidate. 
To our surprise, no study measures such semantic-temporal correlation or examines how well existing methods capture it.

Motivated by the lack of such analysis, we propose an empirical measurement of the ground truth semantic-temporal correlation by mutual information.
On the Amazon dataset, which contains users' reviews on items, we calculate the \textit{mutual information} between history reviews with Top-5 categories at various target-relative positions and the target item with a specific category.
As illustrated in Fig.~\ref{subfig:674}, our results reveal a compelling semantic-temporal pattern.
We then examined the learned correlation of several popular user interest models such as DIN~\cite{din2018}, SASRec~\cite{SASRec} and BST~\cite{BST}, and show the results in Fig.~\ref{fig:ground_truth_and_learned_qc}.
To our surprise, none of these models can capture the semantic-temporal pattern well.

To capture such correlation, the model needs to conduct a 4-way interaction over the quadruplet \texttt{(behavior semantics, target semantics, behavior temporal, target temporal)}. 
Nevertheless, existing user interest methods are not equipped with such quadruple interaction, making them unable to learn semantic-temporal correlation. 
For instance, DIN~\cite{din2018} doesn't consider any temporal information of behaviors.
SASRec~\cite{SASRec} and Bert4Rec~\cite{Bert4Rec} leverage positional embedding to model behavior sequences, but their position is target-agnostic, leading to inadequate modeling of target-aware temporal correlations. 
BST~\cite{BST} conducts self-attention over behaviors and the target, enabling only 3-order explicit interactions.

To this end, we propose a novel model named Temporal Interest Network (TIN), to capture the semantic-temporal correlation.
To achieve this, we adopt Target-aware Temporal Encoding (TTE) to encode the temporal information, which applies to both behaviors and the target itself. 
Specifically, TTE employs the relative position or time interval of each behavior with respect to the target. 
Furthermore, we incorporate Target-aware Attention (TA) and Target-aware Representation (TR), each interacting behaviors with the target. 
Finally, we explicitly incorporate a 4-way interaction by multiplying the output of target-aware attention with that of target-aware representation to capture the quadruple semantic-temporal correlation. 

Each of the three components in TIN is critical in acquiring the semantic-temporal correlation.
To empirically validate the significance of each component, we conduct an ablation study on the Amazon dataset. 
The ablation of each component results in a notable decrease in the GAUC by 0.99\%, 0.73\%, and 0.61\%, respectively.
Notably, all existing methods can be considered ablated variants of TIN, as they lack one or more of these critical components.
The contribution of this paper can be summarized as follows:

\begin{itemize}
    \item We pioneer a study on quantifying the semantic-temporal correlation between behaviors and the target, and reveal the presence of strong semantic-temporal patterns. 
    We examine several popular user interest models and observe that they fail to capture such correlation.
    \item We propose a simple yet effective user interest model named Temporal Interest Network (TIN) to capture semantic-temporal correlation. 
    TIN adopts target-aware temporal encoding and incorporates a 4-way interaction by deploying both target-aware attention and representation.
    \item We conduct comprehensive experiments on two publicly available datasets, and the results demonstrate that \ModelName outperforms state-of-the-art user interest methods. 
\end{itemize}

    \section{Semantic-Temporal Correlation}
\label{sec:temporal_information_matter}

This section presents a novel measurement to quantize the ground truth semantic-temporal correlation between behaviors and the target via \textit{mutual information}.
We then examine several popular user interest models' ability to learn such correlation.

\subsection{Measurement of Ground Truth Semantic-Temporal Correlation}

We investigate the semantic-temporal correlation on the public Amazon dataset~\cite{din2018,dien2019}, which consists of user's reviews on items. 
Regarding temporal correlation, we choose the relative position of each behavior regarding the target without loss of generality.
We choose the category feature of behaviors and the target since its cardinality is intermediate. 
We quantify the ground truth semantic-temporal correlation by the following metric.

\begin{definition}[Category-wise Target-aware Correlation (CTC)]
\label{def:G-TBTC}

CTC is defined as the mutual information~\cite{fwfm2018, shwartz2017opening} between behaviors with category $c_i$ whilst occurring at position $p$: $\mathcal{X}_{C(X)=c_i \land P(X)=p}$ and the user response label~\footnote{The Amazon dataset only contains reviews, \textit{i.e.}, positive responses, on the target item. We generate negative responses by replacing the target item with a random item.} on the target item with category $c_t$: $\mathcal{Y}_{C(Y)=c_t}$. Formally,
\begin{equation}
    \text{Cor} = \text{MI}\left(\mathcal{X}_{C(X)=c_i \land P(X)=p},  \mathcal{Y}_{C(Y)=c_t}\right),
    \label{eq:cor_pi_xt_xi}
\end{equation}
where $C(\cdot)$, $P(\cdot)$ denotes the category or position of the behavior or target. 
For example, the CTC of behaviors with Top-5 categories at various target-relative positions and target with category $c_t = 674$ is illustrated in Fig.~\ref{subfig:674}.
We observe
1) \emph{semantic pattern between matching categories}: behaviors belonging to the same category as the target (the 3rd row, category 674) exhibit a higher degree of correlation compared to other categories. 
2) \emph{temporal decaying pattern}: among the semantically correlated behaviors (\textit{i.e.}, the 3rd row), there is a compelling correlation decrease from the most recent behaviors to the oldest ones. 
\end{definition}



\subsection{How do Existing Methods Capture Semantic-Temporal Pattern?}\label{sec:exist}

We are intrigued by how existing user interest methods capture the observed patterns. 
To investigate this, we select three popular user interest models: DIN~\cite{din2018}, SASRec~\cite{SASRec}, and BST~\cite{BST}, and evaluate the learned correlation of these models on the target category 674, as depicted in Fig.~\ref{subfig:din_plus_674}, Fig.~\ref{subfig:sas_674} and Fig.~\ref{subfig:bst_674}.

We observe notable disparities in the learned correlations of these models compared to the ground truth one. 
DIN fails to capture any temporal pattern, as evidenced by the absence of decay along positions. 
SASRec and BST demonstrate limited proficiency in learning temporal correlation, as there is a slight decay along positions. 
However, they fall short in capturing the semantic correlation, as they learn high correlations for many other categories besides target category 674 (3rd row).
The failure of existing methods to learn semantic-temporal correlation motivates us to develop a more effective user interest model.
For a more comprehensive understanding of how we measure the learned correlation of each model, please refer to Sec.~\ref{subsec:visualization}.

    \section{Temporal Interest Network}
\label{sec:TAR}

\begin{figure*}[htb]
	\centering
	\captionsetup{labelfont=bf}
	\subfigure[TIN Architecture]{
		\begin{minipage}[b]{0.44\linewidth}
			\label{subfig:TIN_whole_architecture}
            \includegraphics[width=1\textwidth]{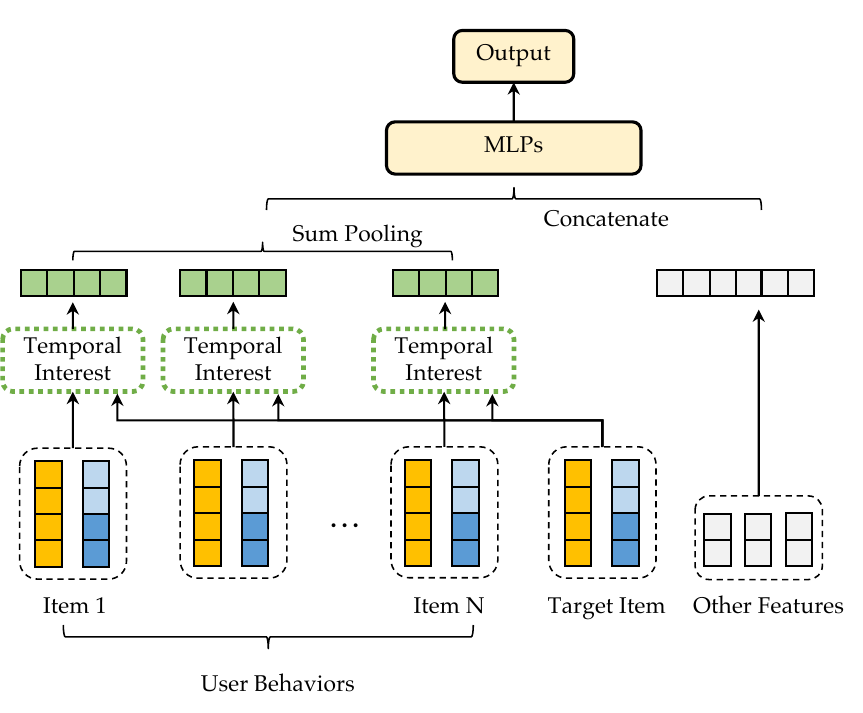}
	\end{minipage}}
	\subfigure[Temporal Interest Module]{
		\begin{minipage}[b]{0.54\linewidth}
			\label{subfig:Temporal_Interest_Module}
			\includegraphics[width=1\textwidth]{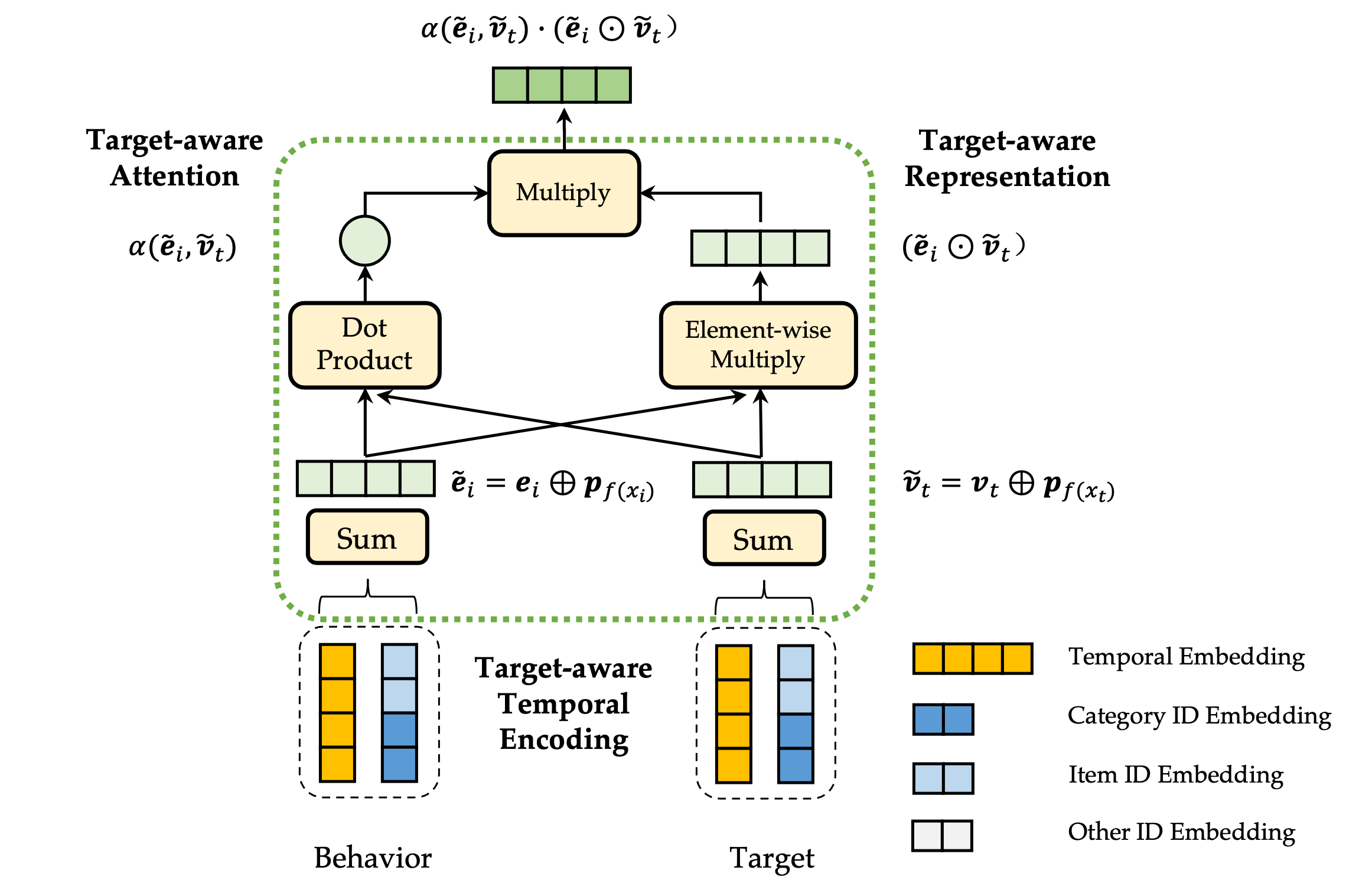}
	\end{minipage}}
	
	\caption{Left: the whole architecture of TIN; Right: the architecture of the Temporal Interest Module, which introduces target-aware temporal encoding, target-aware attention and representation to capture quadruple semantic-temporal correlations between behaviors and the target.}
	\label{fig:network_architecture}
\end{figure*}


Our model adopts the widely used Embedding \& MLP paradigm. 
To begin with, all features from the user side, item side, and context side are transformed into embeddings. 
We then process user behavior features by a Temporal Interest Module (TIM), generating a fixed-length user interest representation.
Subsequently, the output of the TIM and other one-hot encoding features are concatenated and fed into Multi-Layer Perceptrons (MLPs). 
The final output is obtained through a sigmoid function, and the entire model is optimized using the cross-entropy loss.
Refer to Fig.~\ref{fig:network_architecture} for an overview of the model architecture. 

\subsection{Temporal Interest Module}
\label{subsec:temporal_interest_module}

In order to conduct 4-way interaction over the quadruplet \texttt{(behavior semantics, target semantics, behavior temporal, target temporal)} to capture the semantic-temporal correlation, we incorporate the following components.
a) Target-aware Temporal Encoding (TTE): TTE preserves the temporal information of behaviors regarding the target. Combining TTE with the semantic ID embedding, we can capture both semantic and temporal information about behaviors and the target.
b) Target-aware Attention (TA) and c) Target-aware Representation (TR) over the temporally encoded behaviors and target: each conducts a 2-way behavior-target interaction. 
We multiply the output of TA and TR, resulting in an explicit 4-way interaction between behaviors and the target, to capture their semantic and temporal correlation.
Formally,
\begin{equation}
    \bm{u}_{\text{\ModelName}} =  \sum_{i \in \mathcal{H}}  \alpha(\tilde{\bm{e}}_{i}, \tilde{\bm{v}}_{t}) \cdot (\tilde{\bm{e}}_{i} \odot \tilde{\bm{v}}_t)
\end{equation}
where $\tilde{\bm{e}}_i = \bm{e}_i \oplus \bm{p}_{f(X_i)}$ denotes the representation of the $i$-th behavior $X_i$, consisting of both semantic encoding $\bm{e}_i$ and \emph{target-aware temporal encoding} $\bm{p}_{f(i)}$,
similarly, $\tilde{\bm{v}}_t = \bm{v}_t \oplus \bm{p}_{f(X_t)}$ denotes the representation of target item.
$\alpha(\tilde{\bm{e}}_{i}, \tilde{\bm{v}}_{t})$ and  $(\tilde{\bm{e}}_{i} \odot \tilde{\bm{v}}_t)$ denotes the target-aware attention and target-aware representation between the $i$-th behavior and the target, 
$\alpha(\cdot)$ denotes the softmax function, and $\odot$ denotes element-wise multiplication. 
We illustrated the architecture of TIM in Fig.~\ref{subfig:Temporal_Interest_Module}.

\subsubsection{Target-aware Temporal Encoding (TTE)}
\label{subsec:temporal_encoding}

In addition to encoding the ID features of each behavior as embeddings $\bm{e}_i \in \mathcal{R}^d$ to capture semantic meaning, we also incorporate the temporal information of each behavior, focusing on the \emph{target-aware} temporal aspects. 
To achieve this, we employ two distinct methods: TTE based on relative Position and TTE based on Time interval.

\paragraph{Target-aware Temporal Encoding based-on Position (TTE-P)}

Denote the sequence that contains both history clicked items and the target whilst sorted by time as $\{X_1, \dots, X_H, X_t\}$.
TTE-P encodes the temporal information of the $i$-th behavior as $H-i+1$.  
The resulting embedding for this behavior is denoted as $\bm{p}_{f_\text{TTE-P}(X_i)} \in \mathcal{R}^d$. 
Please note that the TTE-P of the target item is assigned a value of 0, representing \emph{the origin}, and its corresponding embedding is denoted as $\bm{p}_{f_\text{TTE-P}(X_t)} = \bm{p}_0$.

\paragraph{Target-aware Temporal Encoding based-on Time interval (TTE-T)}

The TTE-T method encodes the temporal information of the $i$-th behavior by its time interval from the target. 
Let us denote the timestamp of the $i$-th behavior and the target as $TS_i$ and $TS_t$. 
The time interval between them can be represented as  $\tau_i = TS_t - TS_i$.  
To facilitate the learning process, we discretize the time interval by a binning function $b(\cdot)$, such as equal frequency binning. 
Consequently, the TTE-T embedding for the $i$-th behavior is defined as $\bm{p}_{f_\text{TTE-T}(X_i)} = \bm{p}_{b(\tau_i)}$.
Similar to TTE-P, the TTE-T of the target item also represents the origin.

Once we obtain the temporal encoding $\bm{p}_{f(X_i)}$ for a behavior $X_i$,  its final embedding is obtained through element-wise summation with its semantic embedding: $\tilde{\bm{e}}_i = \bm{e}_i \oplus \bm{p}_{f(X_i)}$. Similarly, the target item $X_t$ is represented as $\tilde{\bm{v}}_t = \bm{v}_t \oplus \bm{p}_{f(X_t)}$.
By combining the semantic and temporal embeddings, we get enriched representations that incorporate both the inherent semantic meaning and the temporal context of the behaviors and the target item.

\paragraph{TTE v.s. Chronological Order Encoding}
 
Another widely used temporal encoding is Chronological Order Encoding (COE)~\cite{BST, Bert4Rec, dsin2019, SASRec}, which encodes behaviors based on their chronological order from earliest to oldest, \textit{i.e.}, $f_\text{COE}(X_i) = i$. 
While COE is widely adopted, it has limitations as it is target agnostic. 
It may assign the same weights to behaviors regardless of their relative positions to the target. For instance, consider two behavior sequences with lengths of 10 and 100, respectively. 
According to COE, the first behaviors of both sequences have the same position, \textit{i.e.}, $f_\text{COE}(X_1) = 1$. 
However, their relative positions to the target item differ significantly, \textit{i.e.}, 10 versus 100, indicating varying levels of importance. 
Unfortunately, COE assigns the same position and encoding embedding to both behaviors, disregarding their distinct importance.

There are also other temporal encoding schemes such as Timestamp Encoding (TE)~\cite{DiCycle2022}, which applies an embedding layer to encode a timestamp $t$ containing hour $h$, weekday $w$, month $m$ and etc., into several time embeddings of different granularities, \textit{i.e.}, $f_\text{TE}(X_t) = \text{Emb}(h, w, m, \cdots)$.
While TE can be considered target-aware as long as the target item is incorporated in the attention or representation component, our experiments reveal that it performs worse than TTE. 
We will present a performance comparison of temporal encoding methods in Section~\ref{subsec:temporal_encoding}.

\subsubsection{Target-aware Attention and Target-aware Representation}
\label{subsec:TA}

Building upon target attention methods~\cite{din2018,dien2019,dsin2019}, we aim to capture the importance of each behavior $X_i$ through an attention function.  
In this regard, we adopt the Scaled Dot-Product~\cite{vaswani2017attention,autoattention2022} as our activation function: $
    \alpha(\tilde{\bm{e}}_{i}, \tilde{\bm{v}}_{t}) = \sigma(\frac{\langle\tilde{\bm{v}}_{t},\tilde{\bm{e}}_{i}\rangle}{\sqrt{d}})
$, where $\sigma(z)$ denotes the softmax function on all behaviors of a given user. 

Existing methods typically multiply the attention weight with the representation of each behavior and then concatenate it with the embeddings of the target and other features to be fed into Multi-Layer Perceptrons (MLPs): $g_\text{MLP}([\sum_i \sigma (\bm{e}_i, \bm{v}_t) \bm{e}_i, \bm{v}_t])$. 
However, it has been proved that MLPs struggle to effectively learn dot product or explicit interactions~\cite{revisit2020, zhai2023revisiting}, limiting these methods to conducting only 3-way explicit interactions through the self-attention, \textit{i.e.}, $\sigma (\bm{e}_i, \bm{v}_t) \cdot \bm{e}_i$.

In TIN, each behavior is represented as a second-order representation with respect to the target item, resulting in a target-aware representation: $\tilde{\bm{e}}_i \odot \tilde{\bm{v}}_t$.
Such explicit second-order representation or interaction is widely adopted in existing explicit high-order interaction click-through rate (CTR) models~\cite{fm, ffm2017, fwfm2018, xdeepfm2018, fmfm2021, dcnv22021}.
By multiplying the target-aware attention (TA) with the target-aware representation (TR), TIN explicitly conducts 4-way interactions over the quadruplet \texttt{(behavior semantics, target semantics, behavior temporal, target temporal)}. 
Note that the absence of either TA or TR would cause performance deterioration, which will be discussed in Sec.~\ref{subsec:ablation_study}.

\subsection{Connection to Existing Methods}

The Temporal Information (TI), Target-aware Attention (TA), and Target-aware Representation (TR) have been widely used in existing user interest models for user response prediction~\cite{din2018, SASRec, dien2019, dsin2019, GRU4Rec, Bert4Rec, BST, autoattention2022}. 
In order to analyze the presence of these components, we assign a 3-bit code to represent each model, where each bit indicates the existence (\cmark) or absence (\xmark) of TI, TA, and TR.
It is important to note that TI refers to any form of temporal information, including Temporal/Position Encoding or Recurrent Neural Networks (RNN), and is not limited to Target-aware Temporal Encoding (TTE). 

For instance, the code for \ModelName\ is \cmark\cmark\cmark, indicating the presence of all three components. On the other hand, DIN's code is \xmark\cmark\cmark, as it lacks TI.
Models such as GRU4Rec, SASRec, and BERT4Rec share the same code, \cmark\xmark\cmark, as they incorporate TI and TR but lack TA. 
Conversely, other target attention methods like DIEN, DSIN, and BST have the code \cmark\cmark\xmark, due to the failure of Concat \& MLP to learn dot product~\cite{revisit2020}.
Please refer to Table~\ref{tab:summary_existing_models} for a comprehensive overview of each method's code. 

\subsection{Connections to Transformer}

There are several commonalities between the Temporal Interest Module (TIM) and Scaled Dot-Product Attention proposed in Transformer~\cite{vaswani2017attention}: 
1. they both employ Scaled Dot-Product as the attention function; 
2. they consider both semantic embedding as well as temporal embedding to represent each token or item. 
Using the Transformer terminology, $\tilde{\bm{v}}_t$ and $\tilde{\bm{e}}_i$ in target-aware attention corresponds to \emph{query} (target query) and \emph{key} (behavior key) in Dot-Product Attention, while $\tilde{\bm{e}}_i$ and $\tilde{\bm{v}}_t$ in target-aware representation corresponds to \emph{values} (behavior value and target value).

They differ from each other regarding 1. Scaled Dot-Product Attention is applied between each pair of tokens within the sequence, while TIM is only applied between the target and each item in the sequence. That is, Scaled Dot-Product Attention is a Self-Attention upon a sequence, while TIM is a Target-Attention upon a sequence and a target; 
2. The representation in Scaled Dot-Product Attention is 1st-order only, that is,  $V$, while it's 2nd-order in TIM, that is, $(\tilde{\bm{e}}_{i} \odot \tilde{\bm{v}}_t)$.

    \section{Experiments}
\label{sec:performance_evaluation}

In this section, we aim to address the following research questions (RQ) through experiments  on real-world datasets:

\textbf{RQ1:} How does our proposed Temporal Interest Network (TIN) performance compare to state-of-the-art methods?

\textbf{RQ2:} Is each component of TIN essential for capturing semantic-temporal correlations? What would be the impact if any of these components were removed?

\textbf{RQ3:} How can we effectively measure the learned semantic-temporal correlation of \ModelName\ and other methods? How does \ModelName\ capture such correlations? 

\textbf{RQ4:} How do various temporal encoding methods perform compared to each other?

\begin{table*}[htb]
	\centering
	\captionsetup{labelfont=bf} 
       \addtolength{\tabcolsep}{1pt}
	\caption{Evaluation results on the Alibaba and Amazon dataset. Models with the same background color have the same component code.}
	\label{tab:comp1}
	\begin{tabular}{lc|cccc|cccc}
		\toprule[1.3pt]
		\multirow{2}{*}{Model} & \multirow{2}{*}{Code} & \multicolumn{4}{c}{Amazon} & \multicolumn{4}{|c}{Alibaba}\\
		
		\cmidrule(lr){3-6} \cmidrule(lr){7-10}
		& & Logloss & $\Delta$\% & GAUC & $\Delta$\%  & Logloss & $\Delta$\%  & GAUC & $\Delta$\%  \\
		\midrule
		Avg Pooling \& Concat & \xmark\xmark\xmark  & 0.4908 (1E-3) & - & 0.8445 (2E-3) & - & 0.1969 (1E-3) & - & 0.6074 (5E-4) & -\\
		
		Avg Pooling \& Product & \xmark \xmark \cmark  &  0.4824 (5E-4) & -1.71 &  0.8523 (8E-4) & +0.92 & 0.1960 (1E-3) & -0.46 &  0.6106 (9E-4) & +0.53\\
		
		DIN' & \xmark\cmark\xmark  &  0.4803 (9E-4)  & -2.14 &  0.8536 (4E-4) & +1.08 &  0.1962 (1E-3) & -0.36 &  0.6096 (6E-4) & +0.36 \\ 
		
		
		\rowcolor{LightCyan}
		DIN & \xmark \cmark \cmark   &  \underline{0.4703}  (2E-3) & -4.18 &  0.8590 (1E-3) & +1.72 & 0.1963 (1E-3) & -0.30 & \underline{0.6113} (4E-4) & +0.64 \\
		
		
		\rowcolor{color2}
		GRU4Rec & \cmark\xmark\cmark  &  0.4766 (2E-3) & -2.89 &  0.8574 (2E-3) & +1.53 & 0.1972 (1E-3) & +0.15 &  0.6091 (4E-4) & +0.28\\
		
		\rowcolor{color2}
		SASRec & \cmark\xmark\cmark  &  0.4837 (7E-3) & -1.45 &  0.8497 (4E-3) & +0.62 & \underline{0.1959 (8E-4)} & -0.51 &  0.6091 (4E-4) & +0.28\\

		
		\rowcolor{color2}
		BERT4Rec & \cmark\xmark\cmark  &  0.4833 (5E-3) & -1.53 & 0.8501 (2E-3) & +0.66 & 0.1961 (1E-3) & -0.41 &  0.6096 (9E-4) & +0.36\\

  
		\rowcolor{color3}
		DIEN & \cmark\cmark\xmark  &  0.4807 (8E-3) & -2.06 &  0.8590 (1E-3) & +1.72 & 0.1973 (9E-4) & +0.20 &  0.6108 (6E-4) & +0.56\\
		
		\rowcolor{color3}
		DSIN & \cmark\cmark\xmark  &  0.4726 (2E-3) & -3.71 &  \underline{0.8592} (1E-3) & +1.74 & 0.1964 (2E-3) & -0.25 &  0.6106 (9E-4) & +0.53\\
		
		\rowcolor{color3}
		BST & \cmark\cmark\xmark  &  0.4850 (5E-4) & -1.18 &  0.8500 (9E-4) & +0.65 & \underline{0.1959 (2E-3)} & -0.51 &  0.6096 (6E-4) & +0.36\\

		\midrule
		\ModelName & \cmark\cmark\cmark  &  \textbf{0.4636} (3E-3) & -5.54 &  \textbf{0.8629} (9E-4) & +2.18 & \textbf{0.1954} (2E-3) & -0.76 &  \textbf{0.6144} (4E-4) & +1.15\\
		
		\rowcolor{LightCyan}
		\ModelName w/o TTE & \xmark\cmark\cmark &  0.4752 (2E-3) & -3.18 &  0.8544 (8E-4) & +1.17  & 0.1963 (1E-3) & -0.30 &  0.6135 (7E-4) & +1.00\\
		
		\rowcolor{color2}
		\ModelName w/o TA & \cmark\xmark\cmark  &  0.4758 (3E-3) & -3.06 &  0.8566 (1E-3) & +1.43 & 0.1960 (7E-4) & -0.46 &  0.6094 (8E-4) & +0.33\\

		
		\rowcolor{color3}
		\ModelName w/o TR & \cmark\cmark\xmark  &  0.4743 (2E-3) & -3.36 &  0.8576 (9E-4) & +1.55 & 0.1965 (2E-3) & -0.20 &  0.6127 (1E-3) & +0.87\\
		
  
		\bottomrule[1.3pt]
	\end{tabular}
\end{table*}

\subsection{RQ1: Overall Performance Evaluation}
\label{subsec:performance_evaluation}

We compare \ModelName with the following user interest methods as baselines: Avg Pooling \& Concat~\cite{din2018}, Avg Pooling \& Product~\cite{din2018}, DIN'\footnote[1]{DIN' denotes the architecture described in the paper, consisting of only target-aware attention}, DIN\footnote[2]{DIN denotes the architecture implemented in the released code, which multiplies the attention-weighted pooling of behaviors with the target outside the attention, therefore consists of both target-aware attention and target-aware representation}, GRU4Rec~\cite{GRU4Rec}, SASRec, BERT4Rec~\cite{Bert4Rec}, DIEN~\cite{dien2019}, DSIN~\cite{dsin2019} and BST. Refer to the appendix for the experimental setting.

The performance evaluation results are presented in Table~\ref{tab:comp1}. 
Avg Pooling \& Concat exhibits the poorest performance on both datasets. 
This can be attributed to its lack of all three essential components, rendering it incapable of capturing either temporal correlation or semantic correlation (feature interaction).
In contrast, Avg Pooling \& Product demonstrates a significant performance improvement compared to Sum Pooling \& Concat, with gains of 9.2e-3 and 5.3e-3 on each dataset, respectively. This improvement can be attributed to the inclusion of target-aware representation through the product operation, as discussed in~\cite{revisit2020}.
While DIN' performs worse than Avg Pooling \& Product on three metrics, except for GAUC on the Alibaba dataset, this can be attributed to the limited expressiveness of Temporal Attention (TA) compared to TR.  DIN outperforms DIN' due to its incorporation of target-aware representation.

All the remaining methods, namely GRU4Rec, SASRec, BERT4Rec DIEN, DSIN, and BST, take into account temporal information. 
It is worth noting that many of these methods outperform DIN on the Amazon dataset due to the presence of a strong target-aware temporal correlation, as shown in Fig.~\ref{subfig:674}.
However, on the Alibaba dataset, some of these methods fail to surpass DIN. This discrepancy can potentially be attributed to the dense temporal nature of behaviors in the Alibaba dataset, as discussed in detail in Section~\ref{subsec:temporal_encoding}. The high density of behaviors in relation to time renders the position information less influential or useful in this context.

\ModelName achieves a GAUC of 0.8629 on the Amazon dataset, surpassing the performance of the best-performing baseline (DSIN) by 0.43\%. 
Similarly, on the Alibaba dataset, \ModelName achieves a GAUC of 0.6144, outperforming the best-performing baseline (DIN) by 0.51\%.
Both improvements are statistically significant, demonstrating the superior performance of \ModelName.

\subsection{RQ2: Ablation Study}
\label{subsec:ablation_study}

\subsubsection{Disabling Target-aware Temporal Encoding}
\label{subsubsec:disabling_temporal_encoding}

Disabling TTE in \ModelName renders it incapable of learning the temporal behavior-target correlation, consequently failing to capture the semantic-temporal correlation. 
This deficiency results in a significant performance drop of 9.9e-3 on the Amazon dataset, where a strong temporal behavior-target correlation has been established, as discussed in Section~\ref{sec:temporal_information_matter}.
Similarly, on the Alibaba dataset, the performance experiences a drop of 1.5e-3. This relatively smaller performance drop can be attributed to the narrower time range of behaviors in the Alibaba dataset, which diminishes the strength of the temporal pattern compared to that observed in the Amazon dataset.

\subsubsection{Disabling Target-aware Attention}
\label{subsubsec:disabling_attention_interaction}

In the absence of target-aware attention in TIN, the model regresses to 2-way interaction solely based on the target-aware representation. Furthermore, excluding the attention mechanism results in a target-aware representation that merely pools the semantic and temporal embeddings from all behaviors, thereby losing the temporal information associated with each behavior. To illustrate this, consider a user with only two behaviors, denoted as $i$ and $j$. The resulting representation can be expressed as:

\begin{equation}
    \begin{split}
        \tilde{\bm{e}}_{i} \odot \tilde{\bm{v}}_t
        &= (\bm{v}_i \oplus \bm{p}_{f(X_i)}) \odot (\bm{v}_t \oplus \bm{p}_{f(X_t)}) \\
        &+ (\bm{v}_j \oplus \bm{p}_{f(j)}) \odot (\bm{v}_t \oplus \bm{p}_{f(X_t)}) \\ 
        &= (\bm{v}_i \oplus \bm{p}_{f(X_i)} \oplus \bm{v}_j \oplus \bm{p}_{f(j)}) \odot (\bm{v}_t \oplus \bm{p}_{f(X_t)}).
    \end{split}
 \end{equation}

It is evident that this formulation disrupts the correspondence between each behavior and its respective position. Consequently, this leads to a performance drop of 7.3e-3 and 8.1e-3 on the Amazon and Alibaba datasets, respectively.

\subsubsection{Disabling Target-aware Representation}
\label{subsubsec:disabling_representation_interaction}

Similar to target-aware attention, disabling target-aware representation makes the model degenerate to the 2nd or 3rd order, disabling it to capture semantic-temporal correlation well.  
However, with the presence of position encoding, \ModelName w/o TR is still able to capture the target-aware temporal correlation to some extent within the target-aware attention. 
Therefore, even though there is a 6.1e-3 and 2.7e-3 performance drop on each dataset, \ModelName w/o TR beats \ModelName w/o TA on both datasets and beats \ModelName w/o TTE on the Amazon dataset. 

\subsection{RQ3: Measurement of Learned Correlation}\label{subsec:visualization}

\begin{table*}[!t]
    \centering
    \caption{Measurement of learned semantic-temporal correlation.}
    \begin{tabular}{l|l|l|l}
    \toprule[1.3pt]
        Model & Overall & Attention & Representation  \\
        \midrule
        \ModelName 
        & $e^{\langle \tilde{\bm{e}}_{i}, \tilde{\bm{v}}_{t} \rangle} \cdot \| \tilde{\bm{e}}_{i} \odot \tilde{\bm{v}}_{t} \|_2$ 
        & $e^{\langle \tilde{\bm{e}}_{i}, \tilde{\bm{v}}_{t} \rangle}$ 
        & $ \| \tilde{\bm{e}}_{i} \odot \tilde{\bm{v}}_{t} \|_2$ \\
        
        \midrule
        \ModelName w/o TTE 
        & $e^{\langle \bm{e}_{i}, \bm{v}_{t}  \rangle} \cdot \| \bm{e}_{i} \odot \bm{v}_{t} \|_2$ 
        & $e^{\langle \bm{e}_{i}, \bm{v}_{t}  \rangle} $ 
        & $ \| \bm{e}_{i} \odot \bm{v}_{t} \|_2 $\\
        \midrule 
        \ModelName w/o TA 
        & $\| (\bm{e}_{i} \oplus \bm{p}_{f_\text{TTE-P}(i)}) \odot (\bm{v}_{t} \oplus \bm{p}_{f_\text{TTE-P}(t)}) \|_2$ 
        & $-$ 
        & $  \| \tilde{\bm{e}}_{i} \odot \tilde{\bm{v}}_{t} \|_2 $ \\

        \midrule 
        \ModelName w/o TR 
        & $e^{\langle \bm{e}_{i} \oplus \bm{p}_{f_\text{TTE-P}(i)}, \bm{v}_{t} \oplus \bm{p}_{f_\text{TTE-P}(t)} \rangle} \cdot \| \bm{e}_{i} \oplus \bm{p}_{f_\text{TTE-P}(i)} \|_2$ 
        & $e^{\langle \tilde{\bm{e}}_{i}, \tilde{\bm{v}}_{t} \rangle}$ 
        & $  \| \bm{e}_{i} \oplus \bm{p}_{f_\text{TTE-P}(i)} \|_2$ \\
        \midrule 
        DIN 
        & $ e^{\langle \bm{e}_{i}, \bm{v}_{t}  \rangle} \cdot \| \bm{e}_{i} \odot \bm{v}_{t} \|_2$ 
        & $ e^{\langle \bm{e}_{i}, \bm{v}_{t}  \rangle}$ 
        & $ \| \bm{e}_{i} \odot \bm{v}_{t} \|_2$ \\
        \midrule 
        SASRec 
        & $ \| (\bm{e}_{i} \oplus \bm{p}_{f_{\text{COE}}(i)}) \odot \bm{v}_{t} \|_2 $ 
        & $ - $ 
        & $ \| (\bm{e}_{i} \oplus \bm{p}_{f_{\text{COE}}(i)}) \odot \bm{v}_{t} \|_2 $ \\
        \midrule 
        BST 
        & $ e^{\langle \bm{e}_{i} \oplus \bm{p}_{f_{\text{COE}}(i)}, \bm{v}_{t} \oplus \bm{p}_{f_{\text{COE}}(t)} \rangle} \cdot \| \bm{e}_{i} \oplus \bm{p}_{f_{\text{COE}}(i)} \|_2 $ 
        & $ e^{\langle \bm{e}_{i} \oplus \bm{p}_{f_{\text{COE}}(i)}, \bm{v}_{t} \oplus \bm{p}_{f_{\text{COE}}(t)} \rangle}  $ 
        & $ \| \bm{e}_{i} \oplus \bm{p}_{f_{\text{COE}}(i)} \|_2 $ \\
    \bottomrule[1.3pt]
    \end{tabular}
    
    \label{tab:visualization}
\end{table*}

In this section, we outline how to measure the learned semantic-temporal correlation of various models, including TIN, its three ablated variants, and several representative user interest models. 
The learned semantic-temporal correlation between behavior $X_{i}$ at position $f(X_i)$ and target $X_{t}$ is defined as:

\begin{equation}
    \overline{\text{Cor}}(f(X_i),\bm{e}_{i},\bm{v}_{t}) = e^{z} \cdot \|\bm{r}\|_2.
\end{equation}

where $z$ represents the attention logits and $\bm{r}$ denotes the representation embedding. 
We chose this form since there are connections between the mutual information and the parameterized (intermediate) output of the interaction of two variables~\cite{oord2018representation}.
In our model, the intermediate parameterized output of the interaction before the MLPs is proportional to $e^{z} \cdot \bm{r}$.
For instance, the learned semantic-temporal correlation of \ModelName can be defined as:

\begin{equation}
    \begin{gathered}
        \overline{\text{Cor}}(f(X_i),\bm{e}_{i},\bm{v}_{t})_\text{\ModelName} = \exp^{z_\text{\ModelName}} \cdot \|\bm{r}_\text{\ModelName}\|_2 \\
         = \exp^{\frac{\langle \tilde{\bm{e}}_{i}, \tilde{\bm{v}}_{t} \rangle}{\sqrt{d}}} \cdot \| \tilde{\bm{e}}_{i} \odot \tilde{\bm{v}}_{t} \|_2 
    \end{gathered}
\end{equation}

For the three ablated variants of \ModelName, the corresponding ablated component is disabled in the measurement. In the case of DIN, it possesses both TA and TR but lacks temporal encoding. For SASRec, it lacks TA as it solely applies self-attention over behaviors. The learned semantic-temporal correlation is measured using the representation, which is simplified as the element-wise product between a COE-encoded embedding of the behavior and the target. As for BST, it incorporates TA with COE to capture temporal correlation. However, its representation is target-agnostic since there is no explicit interaction between the behavior and the target.
For a summary of how to measure the learned correlation of all models, please refer to Table~\ref{tab:visualization}.

\begin{figure}[tp!]
	\centering
	\captionsetup{labelfont=bf}
        \vspace{-6pt}
	\subfigure[\ModelName (0.9894)]{
		\begin{minipage}{0.45\linewidth}
			\label{subfig:tui}
   \includegraphics[width=1\linewidth]{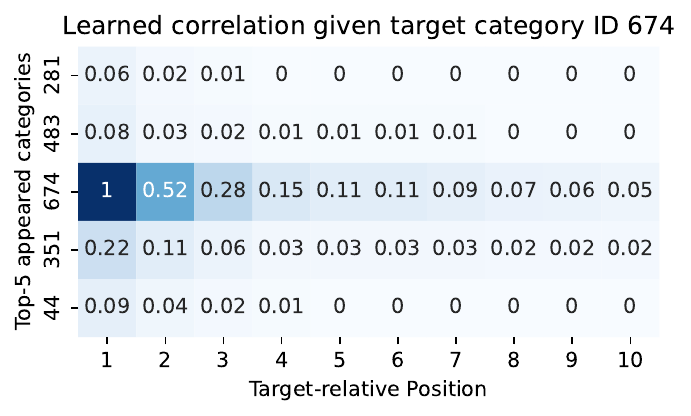}
	\end{minipage}}
	\subfigure[\ModelName w/o TTE (0)]{
		\begin{minipage}{0.45\linewidth}
			\label{subfig:tui_without_pos}
   \includegraphics[width=1\linewidth]{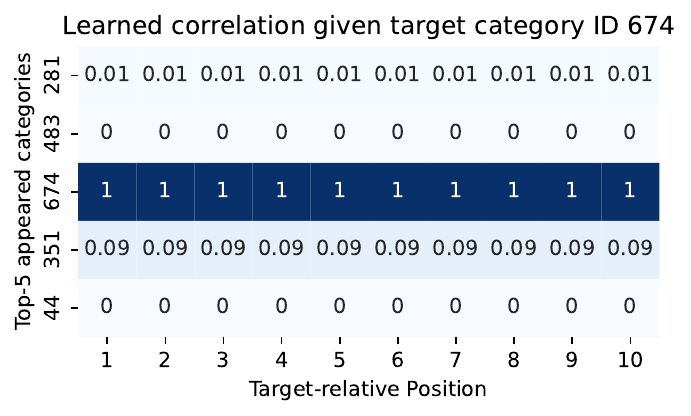}
	\end{minipage}}
	\subfigure[\ModelName w/o TA (-0.6845)]{
		\begin{minipage}{0.45\linewidth}
			\label{subfig:tui_without_attention}
   \includegraphics[width=1\linewidth]{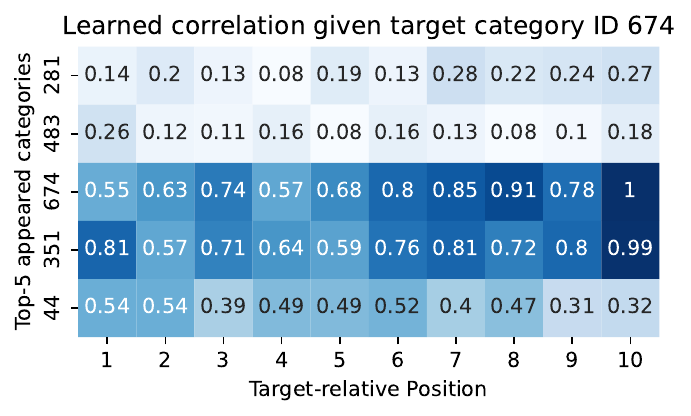}
	\end{minipage}}
	\subfigure[\ModelName w/o TR (0.8179)]{
		\begin{minipage}{0.45\linewidth}
			\label{subfig:tui_without_representation}
   \includegraphics[width=1\linewidth]{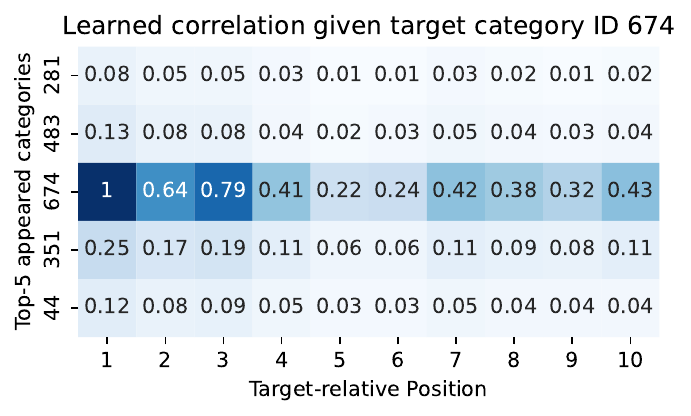}
	\end{minipage}}
 
	\caption{Learned semantic-temporal correlation of TIN and its three ablated variants on the Amazon dataset. The numbers in parentheses represent the Pearson correlation coefficient between the ground truth (as shown in Fig.~\ref{subfig:674}) and the learned correlation on behaviors with the same category as the target (3rd row).}
	\label{fig:general_mi_vs_learned_correlation}
\end{figure}

As depicted in Figure~\ref{fig:general_mi_vs_learned_correlation}, the learned semantic-temporal correlation of \ModelName exhibits a remarkable proximity to the ground truth~\ref{subfig:tui}. 
However, when excluding the Target-aware Temporal Encoding (TTE) component, \ModelName w/o TTE is only capable of learning the semantic correlation between behavior and target categories (as observed in the darkened third row) while failing to capture the underlying temporal pattern. 
Similarly, when excluding the TA component, \ModelName w/o TA not only fails to capture the temporal pattern but also struggles to learn the semantic correlation. 
On the other hand, \ModelName w/o TR demonstrates decent learning of the temporal pattern, but it falls short in capturing the semantic correlation, as evidenced by the non-zero weights assigned to almost all behaviors from other categories.

\begin{table}[bp]
    \centering
    \captionsetup{labelfont=bf}
    \caption{Performance evaluation of temporal encoding methods on the Amazon dataset with various behavior sequence length ranges.}
    \label{tab:comp_tert_p_t}
    \resizebox{0.48\textwidth}{!}{
        \begin{tabular}{l|ccccc}
            \toprule[1.3pt]
    	Range of sequence lengths& $[1,5)$ & $[5,10)$ & $[10,15)$ & $[15,20)$ & $[20,430]$\\
    	\midrule
    	TTE-P & 0.8571 & 0.8630 & 0.8742 & 0.8752 & 0.8728\\
            COE & 0.8564 & 0.8615 & 0.8718  & 0.8731 & 0.8675\\
            \midrule
            $\Delta \%$ (GAUC) & -0.0009 & -0.0017 & -0.0027  & -0.0025 & -0.0061\\
    	\bottomrule[1.3pt]
        \end{tabular}
    }
\end{table}

\begin{table}[bp]
	\centering
	\captionsetup{labelfont=bf}
	\caption{Performance evaluation of various encoding methods.}
	\label{tab:comp_tert}
 \resizebox{0.48\textwidth}{!}{
    	\begin{tabular}{l|cccc|cccc}
    		\toprule[1.3pt]
    		\multirow{2}{*}{}  & \multicolumn{4}{c}{Amazon} & \multicolumn{4}{|c}{Alibaba}\\
    		\cmidrule(lr){2-5} \cmidrule(lr){6-9}
    		& Logloss & $\Delta$\% & GAUC & $\Delta$\% & Logloss & $\Delta$\% & GAUC & $\Delta$\% \\
    		\midrule
    		TTE-P &  0.4636 & - & 0.8629 & - & 0.1954  & - &  0.6144 & -\\
            TTE-T &  0.4484 & -3.28 & 0.8703 & +0.86 & 0.1956 & +0.10 & 0.6139 & -0.08\\
            Timestamp Encoding, PH &  0.4647 & +0.24 & 0.8621 & -0.09 & 0.1960 & +0.31 & 0.6132 & -0.20\\
    		\bottomrule[1.3pt]
    	\end{tabular}
}
\end{table}

\begin{table*}[h]
	\captionsetup{labelfont=bf}
	\caption{A summary of all models w.r.t. the three components: Temporal Information (TI), Target-aware Attention (TA), and Target-aware Representation (TR). COE denotes chronological order encoding, TTE denotes target-aware temporal encoding, $g_{\text{MLP}}(\cdot)$ denotes a Multi-Layer Perceptron and $\left[\cdot\right]$ denotes concatenation. $Q$, $K$, and $V$ represent query, key, and value.}
	\label{tab:summary_existing_models}
	\begin{threeparttable}
	\resizebox{1\textwidth}{!}{
        \begin{tabular}{lcclcclcclclc}
			\toprule[1.3pt]
			Model & \begin{tabular}[c]{@{}l@{}}Temporal\\ Information?\end{tabular} & How & \begin{tabular}[c]{@{}l@{}} TA \end{tabular}  & Attention & \begin{tabular}[c]{@{}l@{}} TR \end{tabular} & Representation & Notes on Representation \\
            \midrule
			Avg Pooling \& Concat  & \xmark & - & \xmark & - & \xmark & $g_\text{MLP}([\bm{e}_i, \bm{v}_t])$ & - \\
   
                Avg Pooling \& Product & \xmark & - & \xmark & - & \cmark & $g_\text{MLP}([\bm{e}_i, \bm{v}_t, \bm{e}_i \odot \bm{v}_t])$ & - \\

			DIN'~\cite{din2018} & \xmark & - & \cmark &  $\sigma(\langle \bm{e}_i, \bm{v}_t \rangle)$ & \xmark &  $g_\text{MLP}([\bm{e}_i, \bm{v}_t])$ & Concat \& MLP's hard to learn dot product~\cite{revisit2020}. \\

                \rowcolor{LightCyan}
			DIN~\cite{din2018} & \xmark & - & \cmark &  $\sigma(\langle \bm{e}_i, \bm{v}_t \rangle)$ & \cmark & $g_\text{MLP}([\bm{e}_i, \bm{v}_t, \bm{e}_i \odot \bm{v}_t])$ &  \begin{tabular}[c]{@{}l@{}} According to code released by the authors. \end{tabular}
			 \\

                SIM~\cite{sim2020} & \cmark & TTE-T & \cmark &  $\sigma(\langle \bm{e}_i, \bm{v}_t \rangle)$ & \xmark &  $g_\text{MLP}([\bm{e}_i, \bm{v}_t])$ & Concat \& MLP's hard to learn dot product~\cite{revisit2020}. \\

                \rowcolor{color2}
			GRU4Rec~\cite{GRU4Rec} & \cmark & GRU & \xmark & -  & \cmark & $\langle \bm{h}_i, \bm{v}_t \rangle$ & \begin{tabular}[c]{@{}l@{}} According to open source code.  \end{tabular} \\

                \rowcolor{color2}
		    SASRec~\cite{SASRec} & \cmark & COE & \xmark & -  & \cmark &   $\langle \tilde{\bm{e}}_i^V, \bm{v}_t\rangle$ & \begin{tabular}[c]{@{}l@{}}According to Eqn. $r_{t,i}=\bm{F}^{(b)}_t \bm{N}_i^T$  of ~\cite{SASRec}. \end{tabular} \\   


                \rowcolor{color2}
			BERT4Rec~\cite{Bert4Rec} & \cmark & COE & \xmark & - & \cmark &  $\langle \tilde{\bm{e}}_i^V, \bm{v}_t\rangle$ & \begin{tabular}[c]{@{}l@{}}According to Eqn.(7) in ~\cite{Bert4Rec}.  \end{tabular} \\


			ExtLayer~\cite{dien2019} & \cmark & AUGRU & \xmark & - & \xmark & - & - \\

                \rowcolor{color3}
			EvoLayer~\cite{dien2019} & \cmark & AUGRU & \cmark & $\sigma( \langle\bm{h}_i, \bm{v}_t \rangle)$ & \xmark & $g_\text{MLP}([\bm{h}_i, \bm{v}_t])$ & - \\

                \rowcolor{color3}
			DSIN~\cite{dsin2019} & \cmark & Bi-LSTM \& COE & \cmark & $\sigma(\langle \tilde{\bm{e}}_i^K, \bm{v}_t^Q \rangle)$  & \xmark & $g_\text{MLP}([\tilde{\bm{e}}_i^V, \bm{v}_t])$ & Concat \& MLP's hard to learn dot product~\cite{revisit2020}. \\

                \rowcolor{color3}
			BST~\cite{BST} & \cmark & COE & \cmark  & $\sigma(\langle \tilde{\bm{e}}_i^K, \tilde{\bm{v}}_t^{Q} \rangle)$ & \xmark  & $g_\text{MLP}([\tilde{\bm{e}}_i^V, \bm{v}_t])$ &  Concat \& MLP's hard to learn dot product~\cite{revisit2020}. \\

                
			\midrule
			
			\ModelName & \cmark & TTE & \cmark & $\sigma(\langle \tilde{\bm{e}}_i, \tilde{\bm{v}}_t \rangle)$ & \cmark &  $g_\text{MLP}([\tilde{\bm{e}}_i, \tilde{\bm{v}}_T, \tilde{\bm{e}}_i\odot\tilde{\bm{v}}_t])$ & - \\

                \rowcolor{LightCyan}
			\ModelName w/o TTE & \xmark & - & \cmark & $\sigma(\langle \tilde{\bm{e}}_i, \tilde{\bm{v}}_t \rangle)$ & \cmark&  $g_\text{MLP}([\tilde{\bm{e}}_i, \tilde{\bm{v}}_t, \tilde{\bm{e}}_i\odot\tilde{\bm{v}}_T])$ & - \\
			
                \rowcolor{color2}
			\ModelName w/o TA & \cmark & TTE & \xmark & - & \cmark&  $g_\text{MLP}([\tilde{\bm{e}}_i, \tilde{\bm{v}}_t, \tilde{\bm{e}}_i\odot\tilde{\bm{v}}_t])$ & - \\
			
                \rowcolor{color3}
			\ModelName w/o TR & \cmark & TTE & \cmark &$\sigma(\langle \tilde{\bm{e}}_i, \tilde{\bm{v}}_t \rangle)$ & \xmark & $g_\text{MLP}([\tilde{\bm{e}}_i, \tilde{\bm{v}}_t])$ & -  \\
			
			\bottomrule[1.3pt]
		\end{tabular}
	}
	\end{threeparttable}
\end{table*}

\subsection{RQ4: Temporal Encoding Methods}
\label{subsec:temporal_encoding}

In \ModelName, we employ TTE-P to encode the target-relative position of historical behaviors. 
This approach differs from the commonly used self-attention position encoding method, Chronological Order Encoding (COE), which encodes behaviors in chronological order from earliest to oldest. 
TTE-P distinguishes itself from COE in that behaviors with the same TTE-P may receive different COE values due to variations in sequence length. Consequently, as the sequence length increases, the disparity between TTE-P and COE becomes more pronounced.

We partitioned the Amazon dataset based on the sequence length to evaluate these two temporal encoding methods and conducted a comparative analysis. 
As shown in Table~\ref{tab:comp_tert_p_t}, TTE-P consistently outperforms COE across all sequence length ranges, with the performance gap widening for longer sequences. Unfortunately, we are unable to perform a similar comparison on the Alibaba dataset since most sequences are padded to a fixed length of 50.

Next, we present the comparison results of TTE-P, TTE-T, and Timestamp Encoding on both datasets in Table~\ref{tab:comp_tert}. 
Specifically, TTE-T split the time interval into 10 buckets with an equal frequency binning function.
Upon replacing TTE-P with TTE-T, we observe an increase in the GAUC by 8.6e-3 on the Amazon dataset and a slight drop of 0.8e-3 on the Alibaba dataset. 
This discrepancy can be attributed to the distinct time ranges of behaviors in the two datasets. 
The Amazon dataset encompasses review data from 1996 to 2014, whereas the Alibaba dataset comprises user behaviors recorded over a continuous period of 8 days. 
The larger time range of the Amazon dataset renders TTE-T more informative than TTE-P. 
Conversely, on the Alibaba dataset with a relatively short time range, TTE-P is more effective in capturing temporal patterns.

\begin{figure}[tp!]
	\centering
	\captionsetup{labelfont=bf}
        \includegraphics[width=1.0\linewidth]{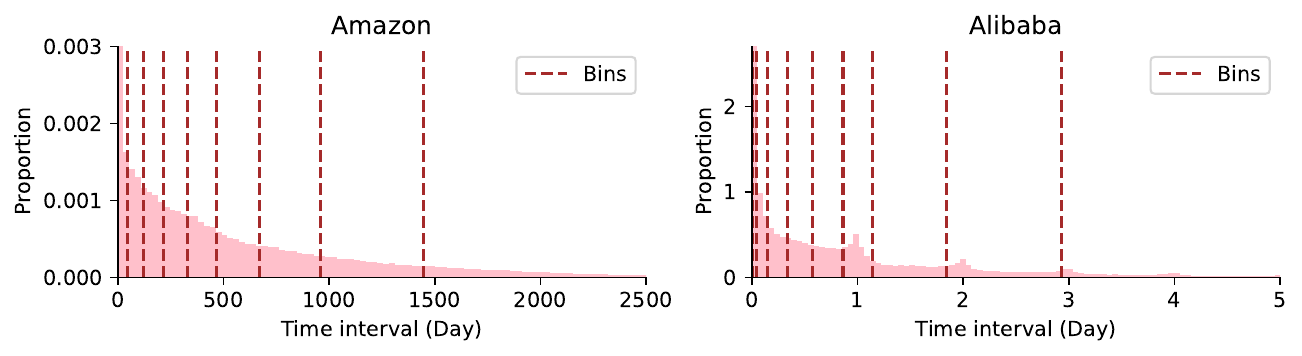}
        \caption{Time interval distribution and binning on Amazon and Alibaba dataset.}
        \label{fig:time_interval}
\end{figure}

\subsection{Existing models regarding the three components}

Table~\ref{tab:summary_existing_models} presents a comprehensive summary of existing models, including \ModelName, with respect to Temporal Information (TI), Target-aware Attention (TA), and Target-aware Representation (TR).
DIN consists solely of TA, while DIN' incorporates both TA and TR. In DIEN~\cite{dien2019}, the Interest Extractor Layer (ExtLayer) and Interest Evolution Layer (EvoLayer) capture TI using the Gated Recurrent Unit (GRU). However, ExtLayer lacks both TA and TR, while EvoLayer lacks TR.
DSIN~\cite{dsin2019} employs the Session Interest Extractor (SIE), which applies self-attention over historical behaviors. The Session Interest Activating (SIA) layer computes an attention weight for each behavior based on the behavior itself and the target item. DSIN then feeds the output of SIA to a Concatenation and Multi-Layer Perceptron (MLP) layer. Consequently, DSIN incorporates TI through self-attention and TA via SIA, but it lacks TR due to the usage of Concatenation and MLP.
BST~\cite{BST} combines both TI (achieved through position encoding) and TA using the Transformer model, which can be viewed as an extension of the DIN model. 
However, BST simply concatenates the output of the Transformer, which represents an attentive pooling of behavior embeddings, with the target item embedding. This concatenated representation then undergoes several MLP layers to make predictions. As mentioned earlier, the Concatenation and MLP are hard to learn the dot product (interaction) effectively~\cite{revisit2020}.

\subsection{Online A/B Testing}

We apply TIN to the user's 2-year clicked ads category sequence feature in Tencent's WeChat Moments and use both TTE-P and TTE-T to capture the temporal correlation.
The backbone model employs a Heterogeneous Experts with Multi-Embedding architecture~\cite{multi-embedding2023, stem2023, pan2024ad}.
Specifically, we learn multiple different feature interaction experts, \textit{e.g.}, GwPFM~\cite{pan2024ad} (a variant of FFM~\cite{ffm2017} and FwFM~\cite{fwfm2018}), IPNN~\cite{pnn2016}, DCN V2~\cite{dcnv22021}, or FlatDNN. 
There are hundreds of features, mostly user behavior features, ads, and context side features.
Multiple embedding tables are learned for all features, each corresponding to one or several experts.
TIN functions as an additional expert alongside the existing three experts, sharing embeddings with GwPFM and FlatDNN.

We conduct online A/B testing from September 2023 to October 2023.
In the baseline, the user interest is modeled by feeding both user behavior features and target ads features to GwPFM or IPNN. 
We previously attempted to use DIN, but it did not surpass the baseline's performance.
This is possible because explicit interaction models such as GwPFM or IPNN can already learn the semantic correlation between behaviors and the target.
During the two-week 20\% A/B testing, TIN demonstrated promising results, achieving a 1.65\% increase in cost and a 1.93\% increase in GMV (Gross Merchandise Value) compared to the baseline.
These improvements were statistically significant according to t-tests. 
Based on our estimation, these performance gains could potentially increase revenue by hundreds of millions of dollars per year.
TIN has been successfully deployed on production in several scenarios.

    \section{Related Work}
\label{sec:related_work}

Deep Interest Network (DIN)~\cite{din2018} introduced the concept of target attention, enabling the learning of attentive weights for each user behavior with respect to a target item. 
Subsequently, several works have emerged following the DIN framework. For instance, DIEN~\cite{dien2019}, DSIN~\cite{dsin2019}, and BST~\cite{BST} focus on modeling interest evolution. 
MIMN~\cite{mimn2019}, SIM~\cite{sim2020}, HPMN~\cite{hpmn2019}, LimaRec~\cite{limarec2021}, and ETA~\cite{eta2021} address the challenge of modeling long sequence interests. 
Additionally, MIND~\cite{mind2019}, ComiRec~\cite{comirec2020}, and LimaRec~\cite{limarec2021} tackle the task of modeling multiple interests.

Another line of research is sequential recommenders~\cite{tang2018personalized,SASRec,yuan2019simple,su2023personalized,hu2022memory,nova2021,difsr-2022,s3-rec2020,cl4srec2022contrastive,MSSR2024}, which aim to capture temporal correlations using either Recurrent Neural Networks (RNN)~\cite{GRU4Rec, RRN, dream2016} or Self-Attention~\cite{SASRec,Bert4Rec,ASReP2021,sse-pt2020}. 
However, these models lack target-aware attention, limiting them to only being able to capture 2nd or 3rd-order interactions and, therefore, unable to learn quadruple semantic-temporal correlation.

    \section{Conclusion}
\label{sec:conclusion}

This paper investigates the crucial semantic-temporal correlation between user behaviors and the target. 
Examination of existing methods reveals that they fail to capture such correlation.
We propose the Temporal Interest Network (TIN) to capture such correlation, which incorporates target-aware temporal encoding, target-aware attention, and target-aware representation.
Comprehensive experiments on two public datasets demonstrate the superiority of TIN over the best-performing baselines.
TIN has been successfully deployed to WeChat Moments on Tencent's advertising platform.

\begin{acks}
\label{sec:acknowledgement}
We gratefully acknowledge the contributions of the following: Xueming Qiu, Lei Mu, Xian Hu, Yaqian Zhang, Ming Yue, Wenbo Liu, Xiaobo Li, Zeen Xu, Jiayu Sun, Xun Liu, Yi Li, Ximei Wang, Junwen Cheng, Yan Tan, Yuxiong Li, Zhaohua Li, Kuo Zhang and Yufei Zheng.

This work was supported by the National Natural Science Foundation of China [U23A20309, 62272302, 62172276, 62372296], 
Shanghai Municipal Science and Technology Major Project [2021SHZDZX-0102], and the Tencent Rhinoceros Project. 
Haolin Zhou, Xinyi Zhou, Xiaofeng Gao and Guihai Chen are in the MoE Key Lab of Artificial Intelligence, Department of Computer Science and Engineering, Shanghai Jiao Tong University, Shanghai, China. Xiaofeng Gao is the corresponding author.
\end{acks}

    \clearpage
    \bibliographystyle{ACM-Reference-Format}
    \bibliography{references}
    \appendix
    \section{Appendix}

\subsection{Experimental Setting}
\label{subsec:experimental_setting}
We assess the effectiveness of TIN by performance evaluation on two popular datasets for user response prediction.

The \textbf{Amazon}\footnote{http://jmcauley.ucsd.edu/data/amazon/} dataset is derived from real product reviews and ratings on the Amazon e-commerce website. 
We conduct experiments on the Electronics subset, which contains 192,403 users, 63,001 goods, 801 categories and 1,689,188 samples. 
Each user in the dataset has a minimum of 5 reviews with various items. 
We adopt the widely employed leave-one-out strategy~\cite{din2018}, that is, we predict the user's feedback for the $n$-th item based on the user's first $n-1$ interaction records. 
The negative samples are generated by replacing the target item in positive samples with a random item from the entire item collection.

The \textbf{Alibaba}\footnote{https://tianchi.aliyun.com/dataset/56} dataset encompasses a comprehensive collection of data obtained from a random selection of 1.14 million users on Taobao's website, resulting in an 8-day advertising display /click log comprising 26 million records.
We partition the dataset as follows: the samples from the first six days (2017-05-06 to 2017-05-11) are allocated for training purposes, the samples occurring on 2017-05-12 are utilized for validation, and the samples from the final day (2017-05-13) are reserved for testing. We follow the data preprocessing procedure of DSIN~\cite{dsin2019}.

For the Amazon dataset, we leverage the \texttt{cate\_id} and \texttt{item\_id} fields to construct the embeddings for both the historical behavior and the target item. Specifically, we compose the embedding $\bm{e}_i$ as follows: $\bm{e}_i = [\bm{c}(X_i), \bm{it}(X_i)] $, where $\bm{c}(X_i)$ denotes the category embedding of $X_i$, $\bm{it}(X_i)$ represents the item embedding of $X_i$, and $[\cdot]$ denotes the concatenation operator. 
On the other hand, for the Alibaba dataset, we utilize the \texttt{cate\_id} and \texttt{brand\_id} fields. In this case, the embedding $\bm{e}_i$ is constructed as follows:  $\bm{e}_i = [\bm{c}(X_i) , \bm{b}(X_i)] $,  where $\bm{b}(X_i)$ denotes the brand embedding of $X_i$. 

For both datasets, we set the embedding dimension $d$ to 64 for all feature embeddings. The hidden layers of the two-layer MLP have dimensions of 80 and 40 for the Amazon dataset, and 200 and 80 for the Alibaba dataset. As the optimizer, we employ Adam~\cite{adam} with a learning rate of 0.001 for the Amazon dataset, and Adagrad~\cite{adagrad} with a learning rate of 0.01 for the Alibaba dataset.
To ensure consistency and comparability, we maintain the remaining settings identical to those used in DIN on the Amazon dataset, and DSIN\footnote{https://github.com/shenweichen/DSIN/tree/master/code} on the Alibaba dataset. We evaluate the performance using Logloss and GAUC.
To account for variability, each experiment is repeated five times, and we report the mean and standard deviation of the results. 

\subsection{Expressiveness Study of Target-aware Attention and Target-aware Representation}
\label{subsec:expressiveness}

Both Target-aware Attention (TA) and Target-aware Representation (TR) are able to learn semantic correlation between behaviors and target. 
However, TR should be more expressive because there exists a solution \emph{by construction} to make it the same as TA by simply summing up all elements of the representation vector of TR. 
We conduct a simple experiment to investigate their expressiveness on DIN which consists of both TA and TR. 
Specifically, we separate the embedding space of TA and TR in DIN, \textit{i.e.}, separate
$\{\bm{v}^\text{TA}, \bm{e}^\text{TA}\} \in \mathbb{R}^{d_\text{TA}}$ from  $\{\bm{v}^\text{TR}, \bm{e}^\text{TR}\} \in \mathbb{R}^{d_\text{TR}}$. 
Formally,

\begin{equation}
    \begin{split}
        \hat{y}_{\text{DIN}} &=  \text{sigmoid}(g_\text{MLP}(\left[\bm{u}_{\text{int}}, \bm{v}_t^{\text{TR}}, \bm{u}_{\text{int}} \odot \bm{v}_t^{\text{TR}}\right])) \\
        \bm{u}_{\text{int}} &=  \sum_{i \in \mathcal{H}}  \alpha_{\text{DIN}}(\bm{e}_{i}^{\text{TA}}, \bm{v}_{t}^{\text{TA}}) \cdot\bm{e}_i^{\text{TR}} \\
        \alpha_{\text{DIN}}(\cdot) &= \sigma(\frac{\left\langle\bm{v}_t^{\text{TA}},\bm{e}_i^{\text{TA}}\right\rangle}{\sqrt{d_\text{TA}}})  \\
    \end{split}
\end{equation}

\begin{figure}
	\centering
	\captionsetup{labelfont=bf}
	\subfigure[Amazon]{
		\begin{minipage}{0.45\linewidth}
			\label{fig:exp_ama}
			\includegraphics[width=1\linewidth]{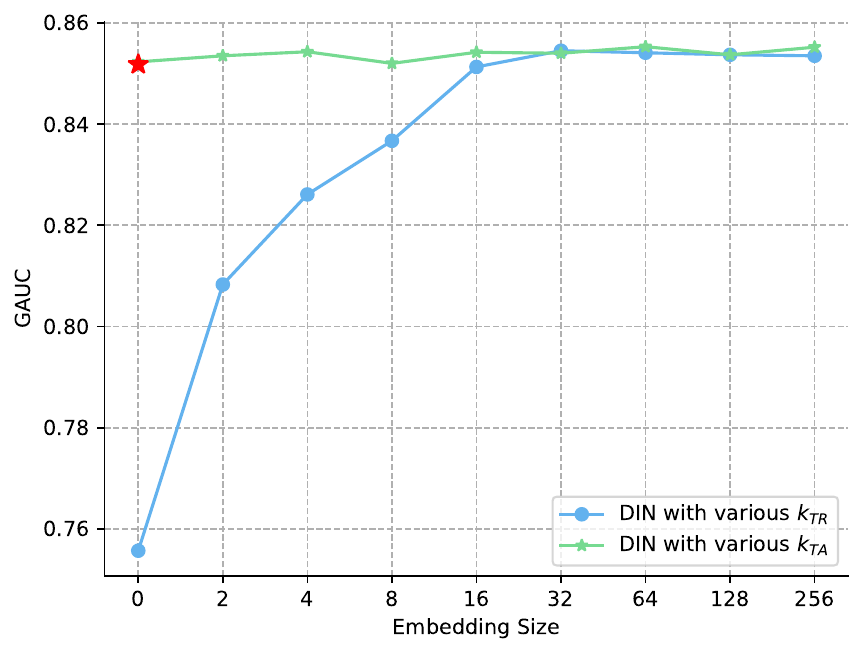}
	\end{minipage}}
	\subfigure[Alibaba]{
		\begin{minipage}{0.45\linewidth}
			\label{fig:exp_ali}
			\includegraphics[width=1\linewidth]{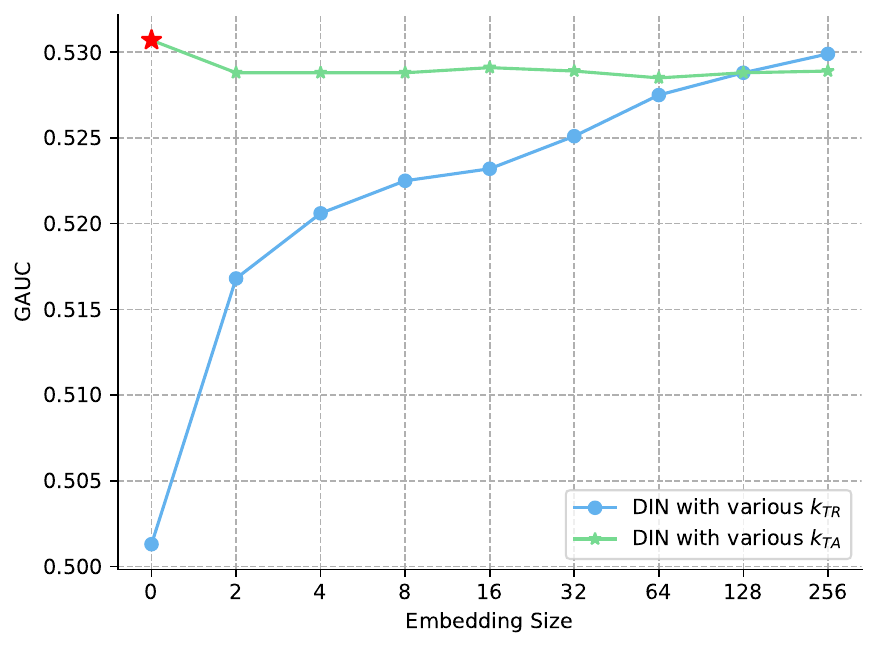}
	\end{minipage}}
	\caption{Expressiveness of target-aware attention and target-aware representation in DIN. The green line denotes the performance of DIN with various attention embedding sizes $d_\text{TA}$, while the blue line denotes the performance of DIN with various representation embedding sizes $d_\text{TR}$. The red star corresponds to totally disabling the target-aware attention.}
	\label{fig:expressiveness}
\end{figure}

We then tune the embedding size of each space, \textit{i.e.}, $d_\text{TA}, d_\text{TR}$ from 0 to 256 in one component while retaining it as 128 for the other one. 
As shown in Fig.~\ref{fig:expressiveness}, the performance drops dramatically with lower embedding size in TR, from 0.854 with $d_\text{TR}=256$ to 0.808 with $d_\text{TR}=2$, shown as the blue line. 
On the other hand, its performance keeps around 0.853 with various embedding sizes ($d_\text{TA}$) from 256 to 2, when retaining $d_\text{TR}$ as 128, shown as the green line. 
This verifies that TR is more expressive than TA in learning the semantic correlation between behaviors and target.

\begin{figure}[!tp]
	\centering
        \vspace{-6pt}
	\captionsetup{labelfont=bf}
	\subfigure[\ModelName (0.8870)]{
		\begin{minipage}[b]{0.31\linewidth}
			\label{subfig:tui_hist}
			\includegraphics[width=1\linewidth]{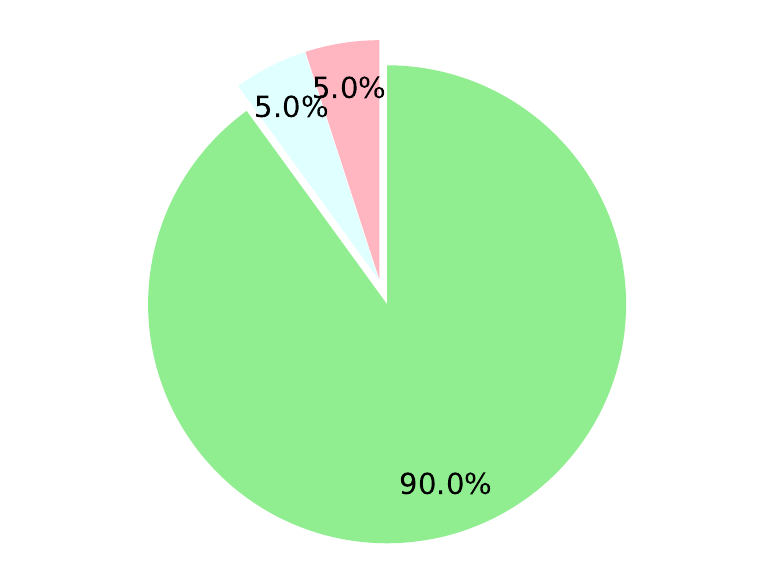}
	\end{minipage}}
	\subfigure[\ModelName w/o TTE (0)]{
		\begin{minipage}[b]{0.31\linewidth}
			\label{subfig:tui_time_hist}
			\includegraphics[width=1\linewidth]{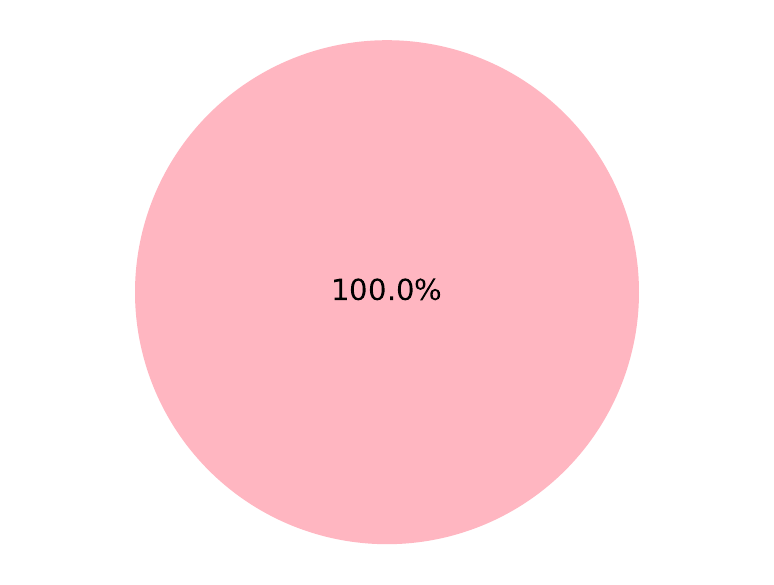}
	\end{minipage}}
	\subfigure[\ModelName w/o TA (0.0712)]{
		\begin{minipage}[b]{0.31\linewidth}
			\label{subfig:bst_hist}
			\includegraphics[width=1\linewidth]{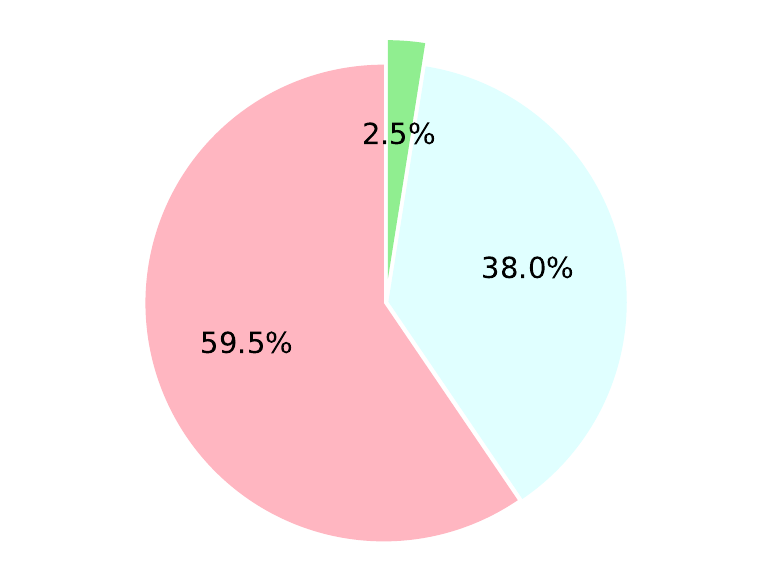}
	\end{minipage}}
		\subfigure[\ModelName w/o TR (0.7142)]{
		\begin{minipage}[b]{0.31\linewidth}
			\label{subfig:tui_hist}
			\includegraphics[width=1\linewidth]{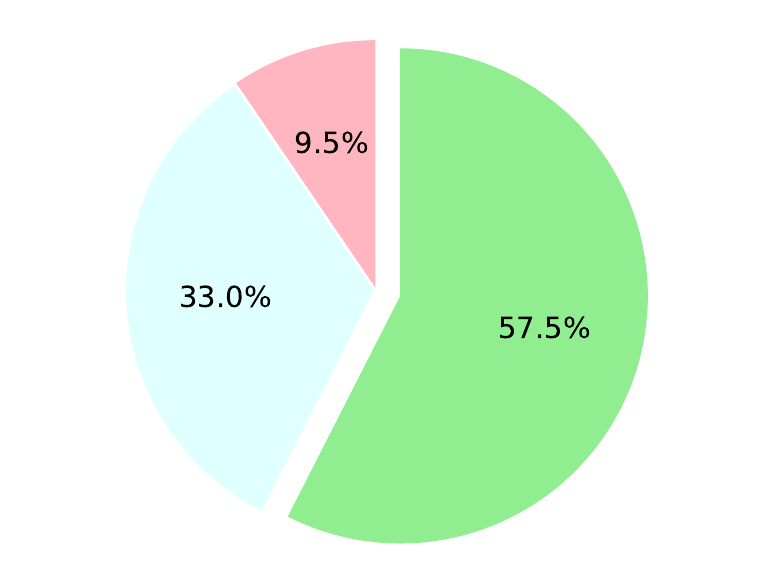}
	\end{minipage}}
        \subfigure[SASRec (0.1062)]{
		\begin{minipage}[b]{0.31\linewidth}
			\label{subfig:bst_hist}
			\includegraphics[width=1\linewidth]{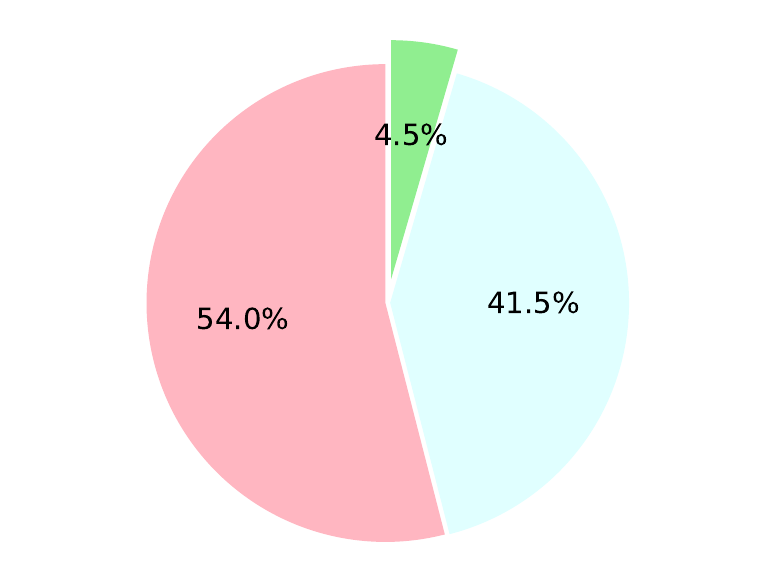}
	\end{minipage}}
		\subfigure[BST (0.3328)]{
		\begin{minipage}[b]{0.31\linewidth}
			\label{subfig:tui_hist}
			\includegraphics[width=1\linewidth]{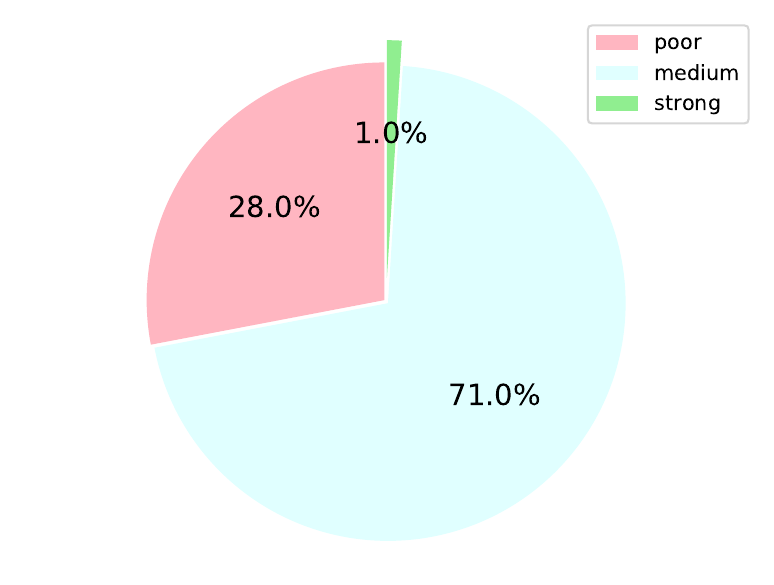}
	\end{minipage}}
	\caption{Distribution of Pearson coefficient regarding ground-truth and learned temporal correlation on the top-200 target categories. The range for poor, medium, and strong coefficients are [-1, 0.2], (0.2, 0.8], and (0.8, 1.0], respectively. The results of DIN are omitted as it's the same with \ModelName w/o TTE.}
	\label{fig:vis_Pearson_update}
\end{figure}






\paragraph{Analysis of semantic-temporal correlation on more categories}

We have also computed the learned semantic-temporal correlation for all the models on the top 200 frequent target categories. To evaluate the correlation between the ground truth and learned Category-wise Target-aware Correlation (CTC), we select the history behaviors with the same category as the target and calculate the Pearson coefficient. The coefficients are then categorized as follows: poor (range of [-1, 0.2]), medium (0.2, 0.8], and strong (0.8, 1.0]. We plot the distribution of these coefficients across the three bins for \ModelName and several other models.
As depicted in Figure~\ref{fig:vis_Pearson_update}, a significant majority (90.0\%) of the categories demonstrate a strong ability to learn the CTC in \ModelName. Similarly, \ModelName w/o TR also exhibits a commendable performance in learning the CTC. However, the remaining models do not achieve comparable results.

\subsection{Visualization on User Behavior Sequences} 

\begin{figure}[bp]
	\centering
	\captionsetup{labelfont=bf}
        \subfigure[The first sample(sequence)]{
		\begin{minipage}{0.45\linewidth}
		\label{subfig:sample_high2}
		\includegraphics[width=1\linewidth]{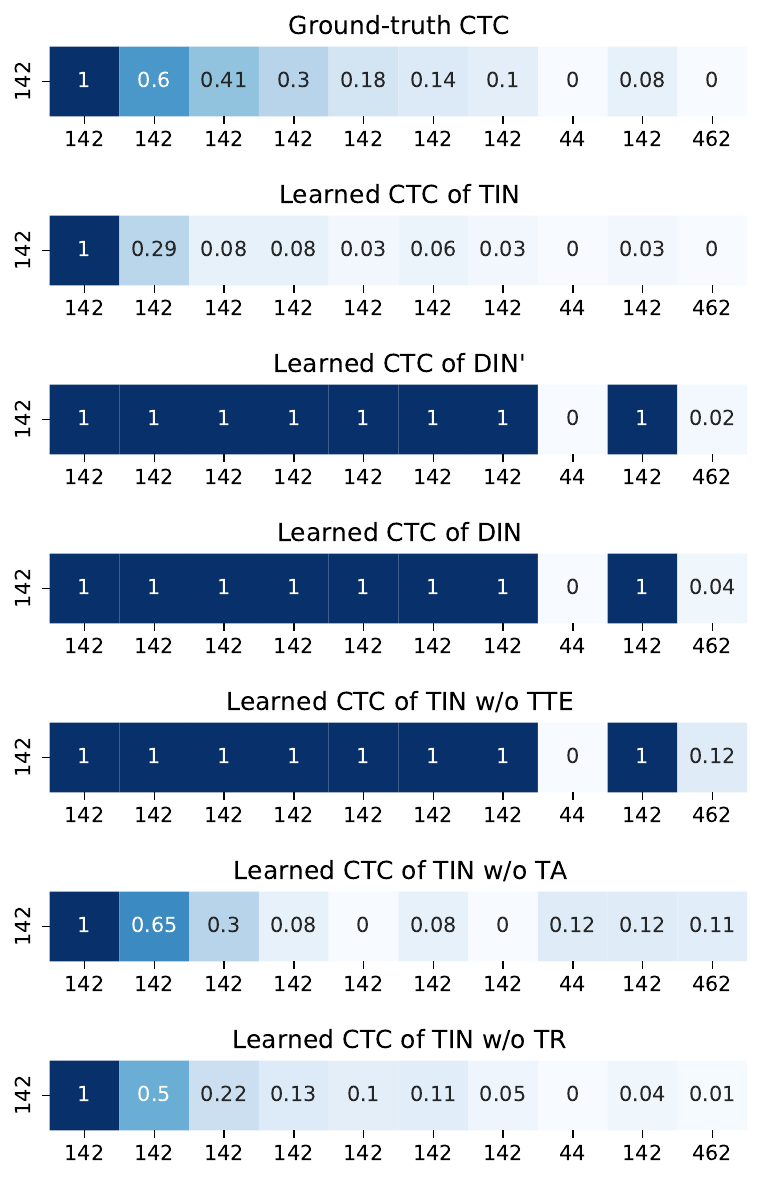}
	\end{minipage}}
	\subfigure[The second sample(sequence)]{
		\begin{minipage}{0.45\linewidth}
			\label{subfig:sample_low}
			\includegraphics[width=1\linewidth]{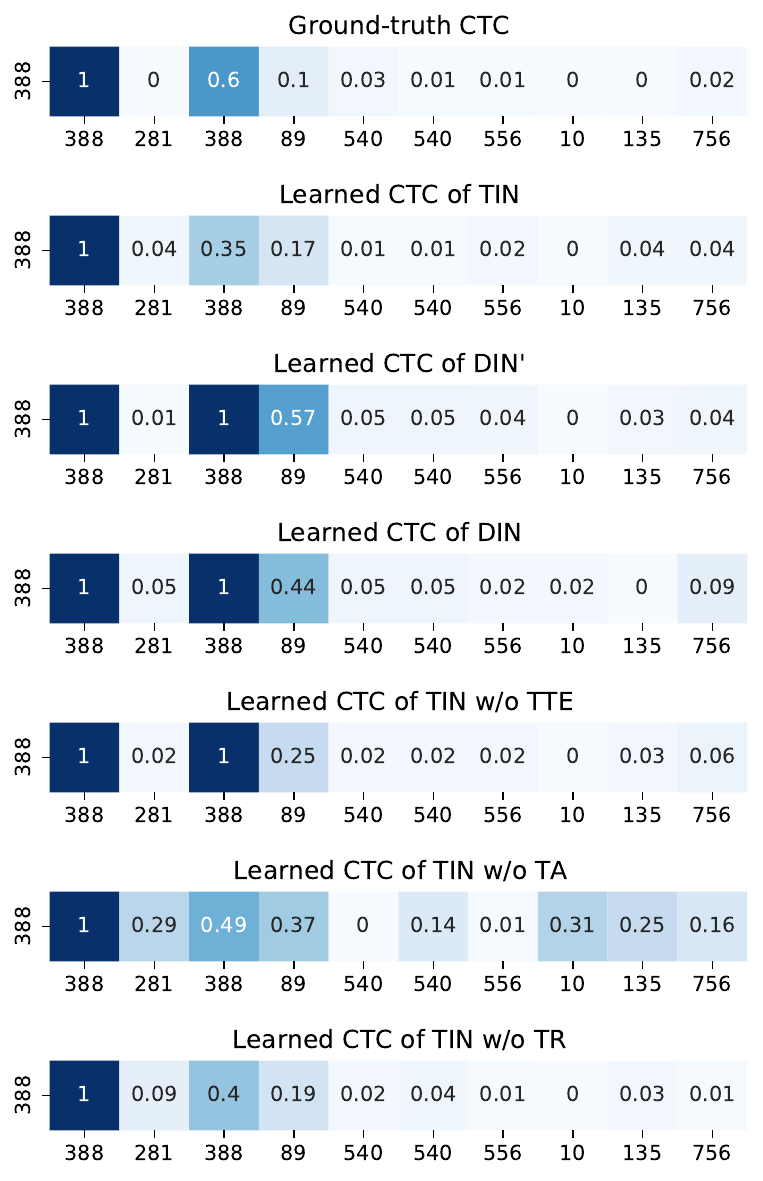}
	\end{minipage}}
	\caption{Visualization of ground-truth CTC and learned CTC between behavior sequence and the target item of 2 random samples from the Amazon dataset.}
    \label{fig:sample}
\end{figure}

We have randomly selected two samples (user behavior sequences) and visualized the ground truth and learned Category-wise Target-aware Correlation (CTC) for the latest 10 behaviors of each sequence in Figure~\ref{fig:sample}.
In the first sequence, 8 out of 10 behaviors share the same category as the target (142). The ground truth CTC exhibits a strong temporally decaying pattern over these 8 behaviors. Notably, both TIN and TIN without Target-aware Representation (TR) successfully capture this correlation, while other models fail to do so. Specifically, DIN', DIN, and \ModelName without Target-aware Temporal Encoding (TTE) learn the same correlation for all 8 behaviors.
In the second sequence, only 2 behaviors have the same category with the target (388), displaying an evident temporal decaying pattern. Conversely, the remaining 8 behaviors belong to diverse categories and has weaker correlation with the target according to the mutual information. 
\ModelName and \ModelName without TR effectively capture both the temporal and semantic correlations. 
However, DIN', DIN, and \ModelName without TTE fail to learn the temporal correlation for the 1st and 3rd behaviors, while \ModelName without Target-aware Attention (TA) fails to capture the weak semantic correlation in the other behaviors.
These visualizations demonstrate the superior performance of \ModelName in capturing both temporal and semantic correlations in specific user behavior sequences.

\end{document}